\newcounter{myctr}
\def\myitem{\refstepcounter{myctr}\bibfont\noindent\ifnum\themyctr>9\else\phantom{0}\fi\hangindent17pt\themyctr.\enskip}

\documentclass{ws-ijqi}

\begin{document}

\markboth{V.I. Man'ko, L.A. Markovich}
{PHOTON DISTRIBUTIONS AND QUADRATURE UNCERTAINTY RELATION $\ldots$}

\catchline{}{}{}{}{}

\title{PHOTON DISTRIBUTIONS AND INFORMATION NONNEGATIVITY
VERSUS QUADRATURE UNCERTAINTY RELATION}

\author{V.I. MAN'KO\footnote{manko@sci.lebedev.ru}}

\address{P.N. Lebedev Physical Institute, Russian Academy of Sciences,\\
Leninskii Prospect 53, Moscow 119991, Russia}

\author{L.A. MARKOVICH\footnote{kimo1@mail.ru}}

\address{Moscow Institute of Physics and Technology, State University,\\ Institutskii Per. 9, Dolgoprudny Moscow Region 141700, Russia,\\V.A. Trapeznikov Institute of Control Sciences, Russian Academy of Sciences, \\Profsoyuznaya 65, 117997 Moscow, Russia.}

\maketitle

\begin{history}
\received{Day Month Year}
\revised{Day Month Year}
\end{history}

\begin{abstract}
Using entropic inequalities for Shannon  entropies new inequalities for some classical polynomials are obtained. To this end,  photon distribution functions for one-, two- and multi-mode squeezed states in terms of Hermite, Laguerre, Legendre polynomials and Gauss' hypergeometric functions are used. The dependence between the violation of the quadrature uncertainty relation, the sign and the existence of the distribution function of such states is considered.
\end{abstract}

\keywords{Hermite polynomials; Gauss' hypergeometric functions; Laguerre polynomials; Legendre polynomials; information inequalities; quantum correlation; quadrature uncertainty relation.}

\section{Introduction}	\label{sec:1}

It is known that photon distributions for one-, two- and multi-mode field states can be represented in terms of special functions.
The photon distribution for the one-mode mixed light with a generic Gaussian Wigner function is presented in terms of Hermite, Laguerre and Legendre  polynomials in Ref.~\refcite{Dodonov}. In Ref.~\refcite{Schleich} the Gauss' hypergeometric functions and the Legendre polynomials are used to represent the photon distribution of the two-mode squeezed states.
\par On the basis of the high popularity of the latter polynomial  representations of the distributions, it is fairly natural to use them  to construct the Shannon entropies. The latter entropies satisfy the inequality called the subadditivity condition \cite{Lieb}. The entropic inequalities for the bipartite systems are used in Ref.~\refcite{Chernega,Mendes} in the framework of the tomographic probability representation of quantum mechanics to characterize two degrees of quantum correlations in the systems. For systems without subsystems the latter inequalities are introduced in Ref.~\refcite{Mar2}. Thus, we can apply the subadditivity condition  in any case, where the set of nonnegative numbers or functions with
the unity sum is arisen. For example, using the latter method the new inequalities for the Jacobi and the Legendre polynomials in case of the system with the spin $j=3/2$ are introduced in Ref.~\refcite{Mar0}. Moreover, for the Lie groups like $SU(2)$ and $SU(1,1)$ the unitary irreducible representations are well known. Hence, they can be represented  in terms of the Jacobi, the Legendre, the Laguerre and the Gauss' hypergeometric polynomials, etc. New inequalities for such special functions  as the Jacobi and Gauss' hypergeometric polynomials are obtained in Ref.~\refcite{Mar10}.
\par The aim of our paper is to derive new relations between classical polynimials including the Hermite, the Legendre, the Laguerre polynimials and the Gauss' hypergeometric functions. To this end, we consider the special polynomial representation of the photon distributions for the Gaussian states and the invertible mapping method \cite{Ibort}. We investigate the dependence between the violation of the quadrature uncertainty relation, the sign and the existence of the photon distribution function. It is shown that the violation of the
quadrature uncertainty relation leads to the negative or complex values of the distribution function.
In other words, the function cannot be longer a probability.
\par The paper is organized as follows. In Sec. ~\ref{sec:2} we recall results known for the one- and the two-mode squeezed states. The probability distributions for such states are written in terms of the several polynomials. In Sec. ~\ref{sec:3} the new inequalities for the Hermite, the Laguerre and the Legendre polynomials are written. In Sec.~\ref{sec:4} we focus on the study of the connection between the quadrature uncertainty relation and the existence of the photon probability distribution. The obtained results are illustrated on examples of a coherent state, a nonlinear coherent state \cite{Lopez} and a squeezed and correlated state \cite{Marian,Marian2} in Sec.~\ref{sec:5}.

\section{Mixed light with the Gaussian Wigner Function}\label{sec:2}

\subsection{The Probability Function of the One-mode Mixed Light}
    Let us have a quantum state with a density matrix $\widehat{\rho}$ described by a generic gaussian Wigner function  $W(p,q)$. It is known that the latter function depends on five parameters. The first two are the mean values of the momentum and position \cite{Dodonov}
\begin{eqnarray*}<q>&=&Tr \widehat{\rho}\widehat{q},\quad
<p>=Tr \widehat{\rho}\widehat{p}.
 \end{eqnarray*}
 The operators $\widehat{q}$ and $\widehat{p}$ are defined by means of the photon creation $\widehat{a}^{\dagger}$ and annihilation $\widehat{a}$ operators as
 \begin{eqnarray*}\widehat{q}&=&\frac{\widehat{a}+\widehat{a}^{\dagger}}{\sqrt{2}},\quad
\widehat{p}=\frac{\widehat{a}-\widehat{a}^{\dagger}}{i\sqrt{2}}.
 \end{eqnarray*}
 Other three parameters are the matrix elements of the real and symmetric dispersion (covariance) matrix
 \begin{eqnarray}\label{34}\Sigma&=&\left(
                             \begin{array}{cc}
                               \sigma_{pp} & \sigma_{pq} \\
                               \sigma_{pq} & \sigma_{qq} \\
                             \end{array}
                           \right),
  \end{eqnarray}
  which are defined as follows
   \begin{eqnarray*}\sigma_{pp}&=&Tr \widehat{\rho}\widehat{p}^2-<p>^2,\quad
   \sigma_{qq}=Tr \widehat{\rho}\widehat{q}^2-<q>^2,\\
   \sigma_{pq}&=&\frac{1}{2}Tr \widehat{\rho}(\widehat{p}\widehat{q}+\widehat{q}\widehat{p})-<p><q>.
  \end{eqnarray*}
 Hence, the Gaussian Wigner function can be written as
\begin{eqnarray*}W(p,q)&=&\det{\Sigma}^{-1/2}\exp\left(-\frac{\sigma_{qq}\overline{p}+\sigma_{pp}\overline{q}-2\sigma_{pq}\overline{p}\overline{q}}{2\det{\Sigma}}\right).
 \end{eqnarray*}
 The uncertainty relations among canonical operators impose the constraint \cite{Olivares} on the dispersion matrix \eqref{34} that reads as
  \begin{eqnarray*}\Sigma+\frac{i}{2}\left(
                                       \begin{array}{cc}
                                         0 & 1 \\
                                         -1 & 0 \\
                                       \end{array}
                                     \right)\geq0.
  \end{eqnarray*}
  Next, the probability to have $n$ photons in the state with the density operator $\widehat{\rho}$ is
  \begin{eqnarray*}P_n=Tr \widehat{\rho}|n><n|,\quad n=0,1,2,\ldots,
  \end{eqnarray*}
  where $\widehat{a}^{\dagger}\widehat{a}|n>=n|n>$. In Ref.~\refcite{Dodonov} the latter probability is expressed in terms of the Hermite polynomials as
   \begin{eqnarray}\label{1}P_n=P_0\frac{H_{nn}^{\{R\}}(y_1,y_2)}{n!},
  \end{eqnarray}
  where
  \begin{eqnarray*}P_0&=&\left(\det{\Sigma}+\frac{Tr \Sigma}{2}+\frac{1}{4}\right)^{-1}\\
  &\cdot&\exp{\frac{-<p>^2\left(\sigma_{qq}+\frac{1}{2}\right)-<q>^2\left(\sigma_{pp}+\frac{1}{2}\right)+2\sigma_{pq}<q><p>}{Tr \Sigma+2 \det{\Sigma}+\frac{1}{2}}}
  \end{eqnarray*}
  is the probability to have no photon.
  The matrix $R$ determining the Hermite polynomials has the following elements
   \begin{eqnarray*}&&R_{11}=R_{22}^{*}=\frac{\sigma_{pp}-\sigma_{qq}-2i\sigma_{pq}}{Tr \Sigma+2 \det{\Sigma}+\frac{1}{2}},\quad
   R_{12}=\frac{\frac{1}{2}-\det{\Sigma}}{Tr \Sigma+2 \det{\Sigma}+\frac{1}{2}}
  \end{eqnarray*}
  and the arguments of the polynomial are of the form
  \begin{eqnarray*}y_1=y_2^{*}=\frac{(Tr \Sigma-1)<z^{*}>+(\sigma_{pp}-\sigma_{qq}+2i\sigma_{pq})<z>}{Tr \Sigma-2 \det{\Sigma}-\frac{1}{2}},<z>=\frac{<q>+i<p>}{\sqrt{2}}.
  \end{eqnarray*}
  Using the sum rule for the Hermite polynomials
  \begin{eqnarray*}H_{nn}^{\{R\}}(y_1,y_2)=n!^2\left(\frac{R_{11}R_{22}}{4}\right)^{\frac{n}{2}}\sum\limits_{k=0}^{n}
  \frac{\left(-\frac{2R_{12}}{\sqrt{R_{11}R_{22}}}\right)^{k}}{(n-k)!^2k!}H_{n-k}\left(z_1\right)
  H_{n-k}\left(z_2\right),
   \end{eqnarray*}
   \begin{eqnarray*}z_1=\frac{R_{11}y_1+R_{12}y_2}{2\sqrt{R_{11}}},\quad
  z_2=\frac{R_{12}y_1+R_{22}y_2}{2\sqrt{R_{22}}}
   \end{eqnarray*}
  the photon distribution function \eqref{1} can be rewritten as
     \begin{eqnarray*}P_n&=&P_0n!\left(\frac{R_{11}R_{22}}{4}\right)^{\frac{n}{2}}\sum\limits_{s=0}^{n}\frac{\left(-\frac{2R_{12}}{\sqrt{R_{11}R_{22}}}\right)^{s}}{(n-s)!^{2}s!}
     \Bigg|H_{n-s}\left(z_1\right)\Bigg|^{2}.
  \end{eqnarray*}
  Also, the latter distribution function can be rewritten in terms of the Laguerre polynomials $\frac{1}{\sqrt{1-z}}\exp\left(\frac{xz}{z-1}\right)=\sum_{n=0}^{\infty}z^nL_n^{-\frac{1}{2}}(x)$  as follows
         \begin{eqnarray}\label{5}P_n&=&\sum\limits_{s=0}^{n}D(n,s)L_s^{-\frac{1}{2}}(x_1)L_{n-s}^{-\frac{1}{2}}(x_2),\\\nonumber
         D(n,s)&\equiv& P_0(-1)^n(R_{12}-\sqrt{R_{11}R_{22}})^{s}(R_{12}+\sqrt{R_{11}R_{22}})^{n-s}.
  \end{eqnarray}
  The numbers $x_1$ and $x_2$ are given by the following expressions
         \begin{eqnarray*}x_{1,2}&=&\frac{1}{4}(R_{12}\mp\sqrt{R_{11}R_{22}})\Bigg(2(R_{11}y_1+R_{12}y_2)(R_{12}y_1+R_{22}y_2)\\
         &\mp&\sqrt{\frac{R_{11}}{R_{22}}}(R_{11}y_1+R_{12}y_2)^2\mp\sqrt{\frac{R_{22}}{R_{11}}}(R_{12}y_1+R_{22}y_2)^2\Bigg).
  \end{eqnarray*}
Hereafter, the probabilities expressed in terms of the Hermite and the Laguerre polynomials will be used in the entropic and the information inequalities.

\subsection{The Probability Function of the Two-mode Squeezed Light}
     Let us have a system of $n$ photons in both modes corresponding to the surface of the four-dimensional sphere $(p_1^2+x_1^2+p_2^2+x_2^2)/2=n$, centered at the origin of the four-dimensional phase space. In Ref.~\refcite{Schleich} the probability of counting $n=2k$ photons for the case of two independently squeezed oscillators reads
   \begin{eqnarray}\label{4}P_{2k}(s_1,s_2)&=&\sqrt{1-s_1}\sqrt{1-s_2}s_2^k{}_2F_{1}(-k,1/2,1;1-s_1/s_2),
  \end{eqnarray}
  where ${}_2F_{1}$ denotes the Gauss' hypergeometric function, $s_j=\tanh{r_j}^2$ and $r_j$, $j=1,2$ are the two squeezing parameters.
  The latter probability can be represented in terms of the associated Legendre polynomials $\mathcal{L}_l^k$ as
     \begin{eqnarray}\label{3}P(n_1,n_2)&=&N\exp\left(-\Bigg|\ln(\frac{n_1!}{n_2!})\Bigg|\right)F_1^{\frac{n_1-n_2}{2}}F_2^{\frac{n_1+n_2}{2}}
     \Bigg|\mathcal{L}_{\frac{|n_1+n_2|}{2}}^{\frac{|n_1-n_2|}{2}}(F_3)\Bigg|^2,
  \end{eqnarray}
  where we use the notations
       \begin{eqnarray*}N&=&\frac{8\sqrt{(A_1B_1-C_1^2)}}{|(2A+1)(2B+1)-4C^2|}\sqrt{(A_1B_1-C_1^2)},\\
     F_1&=&\Bigg|\frac{4AB+2A-2B-1-4C^2}{4AB-2A+2B-1-4C^2}\Bigg|,\quad  F_2= \Bigg|\frac{4AB-2A-2B+1-4C^2}{4AB+2A+2B+1-4C^2} \Bigg|,\\
     F_3&=&\frac{-4C}{\sqrt{4AB+2A+2B+1-4C^2}\sqrt{4C^2-4AB+2A+2B-1}}.
  \end{eqnarray*}
  The $A,A_1,B,B_1,C$ are defined in Ref.~\refcite{Schleich}.
 Hence, we have the expressions for the distribution function of the squeezed light in terms of the Legendre polynomials. The latter expressions will be used in the next section to write new inequalities for the special functions.

\section{Inequalities for the Hermite, the Laguerre and the Legendre Polynomials}\label{sec:3}

 In spirit of Ref.~\refcite{Mar0,Mar10} let us introduce the matrix
 \begin{eqnarray}\label{13}
\rho_{12}&=&\left(
              \begin{array}{cccc}
                P_0 & 0 & 0 & \cdots \\
                0& P_1& 0& \cdots \\
                0 & 0 & P_2 & \cdots \\
                \vdots & \vdots& \vdots & \ddots \\
              \end{array}
            \right)
\end{eqnarray}
with the diagonal elements $P_n$ defined in \eqref{1}. We partition the latter matrix into the block matrices of the size $2\times 2$. Hence, the two new matrices can be constructed as follows
 \begin{eqnarray*}
\rho_{1}&=&\left(
              \begin{array}{cccc}
                P_0 + P_1& 0 & 0 & \cdots \\
                0& P_2 +P_3& 0& \cdots \\
                0 & 0 & P_4 +P_5& \cdots \\
                \vdots & \vdots& \vdots & \ddots \\
              \end{array}
            \right),
\end{eqnarray*}
 \begin{eqnarray*}
\rho_{2}&=&\left(
            \begin{array}{cc}
              P_0  & 0 \\
              0 & P_1 \\
            \end{array}
          \right)+\left(
            \begin{array}{cc}
              P_2  & 0 \\
              0 & P_3 \\
            \end{array}
          \right)+\ldots
=
\left(
  \begin{array}{cc}
    P_0 + P_2+P_4 +\ldots& 0 \\
    0 & P_1 + P_3+P_5 \ldots \\
  \end{array}
\right).
\end{eqnarray*}
Next, we consider the Shannon entropies \cite{Shannon} of the whole system and its subsystems
\begin{eqnarray}\label{9}\mathcal{H}(12)&=&-\sum\limits_{k=0}^{\infty}P_k\ln P_k,\quad
\mathcal{H}(1)=-\sum\limits_{k=0}^{\infty}(P_{2k}+P_{2k+1})\ln (P_{2k}+P_{2k+1}),\\\nonumber
\mathcal{H}(2)&=&-\left(\sum\limits_{k=0}^{\infty}P_{2k}\right)\ln \left(\sum\limits_{k=0}^{\infty}P_{2k}\right)
-\left(\sum\limits_{k=0}^{\infty}P_{2k+1}\right)\ln \left(\sum\limits_{k=0}^{\infty}P_{2k+1}\right).
\end{eqnarray}
The Shannon information is $I=\mathcal{H}(1)+\mathcal{H}(2)-\mathcal{H}(12)$ and  the subadditivity condition reads as
\begin{eqnarray}\label{10}\mathcal{H}(1)+\mathcal{H}(2)\geq\mathcal{H}(12).
\end{eqnarray}
Thus, using the entropies \eqref{9} we can rewrite the subadditivity condition \eqref{10} as
\begin{eqnarray}\label{27}&-&\left(\sum\limits_{k=0}^{\infty}P_{2k}\right)\ln \left(\sum\limits_{k=0}^{\infty}P_{2k}\right)-
-\left(\sum\limits_{k=0}^{\infty}P_{2k+1}\right)\ln \left(\sum\limits_{k=0}^{\infty}P_{2k+1}\right)\\\nonumber
&-&\sum\limits_{k=0}^{\infty}(P_{2k}+P_{2k+1})\ln (P_{2k}+P_{2k+1})
\geq -\sum\limits_{k=0}^{\infty}P_k\ln P_k.
\end{eqnarray}
If we substitute the Hermite polynomial representation of the probability function \eqref{1} in \eqref{27}  we can write the new inequality for the Hermite polynomials
\begin{eqnarray}\label{2}
&-&\left(\sum\limits_{k=0}^{\infty}\mathrm{H}_{2k+1,2k+1}\right)\ln \left(P_0\sum\limits_{k=0}^{\infty}\mathrm{H}_{2k+1,2k+1}\right)\\
&-&\sum\limits_{k=0}^{\infty}\left(\mathrm{H}_{2k,2k}+\mathrm{H}_{2k+1,2k+1}\right)\ln  \left( P_0\left(\mathrm{H}_{2k,2k}+\mathrm{H}_{2k+1,2k+1}\right)\right)\nonumber\\
&-&\left(\sum\limits_{k=0}^{\infty}\mathrm{H}_{2k,2k}\right)\ln \left(P_0\sum\limits_{k=0}^{\infty}\mathrm{H}_{2k,2k}\right)\geq-\sum\limits_{k=0}^{\infty}\mathrm{H}_{k,k}\ln \left(P_0\mathrm{H}_{k,k}\right),\nonumber
\end{eqnarray}
where we use the notation $\mathrm{H}_{k,k}\equiv\frac{H_{k,k}^{\{R\}}(y_1,y_2)}{k!}$.
What's more, using the representation \eqref{5} the new inequality for the Laguerre polynomials can be written as
\begin{eqnarray}\label{28}
&-&\left(\sum\limits_{k=0}^{\infty}\sum\limits_{s=0}^{2k+1}\mathrm{L}_{2k+1,s}\right)\ln \left(\sum\limits_{k=0}^{\infty}\sum\limits_{s=0}^{2k+1}\mathrm{L}_{2k+1,s}\right)\\\nonumber
&-&\sum\limits_{k=0}^{\infty}\left(\sum\limits_{s=0}^{2k}\mathrm{L}_{2k,s}+\sum\limits_{s=0}^{2k+1}\mathrm{L}_{2k+1,s}\right)\ln \left(\sum\limits_{s=0}^{2k}\mathrm{L}_{2k,s}+\sum\limits_{s=0}^{2k+1}\mathrm{L}_{2k+1,s}\right)\\\nonumber
&-&\left(\sum\limits_{k=0}^{\infty}\sum\limits_{s=0}^{2k}\mathrm{L}_{2k,s}\right)\ln \left(\sum\limits_{k=0}^{\infty}\sum\limits_{s=0}^{2k}\mathrm{L}_{2k,s}\right)\geq -\sum\limits_{k=0}^{\infty}\sum\limits_{s=0}^{k}\mathrm{L}_{k,s}\ln \sum\limits_{s=0}^{k}\mathrm{L}_{k,s},
\end{eqnarray}
where $\mathrm{L}_{2k,s}\equiv D(2k,s)L_s^{-\frac{1}{2}}(x_1)L_{2k-s}^{-\frac{1}{2}}(x_2)$.
\par To obtain inequalities for the Legendre polynomials let us write the Shannon entropy of the whole system as
\begin{eqnarray*}\mathcal{H}(12)&=&-\sum\limits_{n_1,n_2=0}^{\infty}P(n_1,n_2)\ln P(n_1,n_2),
\end{eqnarray*}
where the probabilities $P(n_1,n_2)$ are defined by \eqref{3}.
The Shannon entropies for the subsystems are the following
\begin{eqnarray*}\mathcal{H}(1)&=&-\sum\limits_{n_1=0}^{\infty}P(n_1,n_2)\ln P(n_1,n_2),\quad
\mathcal{H}(2)=-\sum\limits_{n_2=0}^{\infty}P(n_1,n_2)\ln P(n_1,n_2).
\end{eqnarray*}
Hence, the subadditivity condition \eqref{10} can be rewritten as
   \begin{eqnarray*}&-&\sum\limits_{n_1=0}^{\infty}P(n_1,n_2)\ln P(n_1,n_2)-\sum\limits_{n_2=0}^{\infty}P(n_1,n_2)\ln P(n_1,n_2)\\
   &\geq& -\sum\limits_{n_1,n_2=0}^{\infty}P(n_1,n_2)\ln P(n_1,n_2).
  \end{eqnarray*}
  Using the  representation \eqref{3} in terms of the  Legendre polynomials  we can rewrite the latter inequality as
   \begin{eqnarray}\label{11}&-&\sum\limits_{n_1=0}^{\infty}T(n_1,n_2)
     \Bigg|\mathcal{L}_{\frac{|n_1+n_2|}{2}}^{\frac{|n_1-n_2|}{2}}(F_3)\Bigg|^2\ln \left(NT(n_1,n_2)
    \Bigg|\mathcal{L}_{\frac{|n_1+n_2|}{2}}^{\frac{|n_1-n_2|}{2}}(F_3)\Bigg|^2\right)\\\nonumber
    &-&\sum\limits_{n_2=0}^{\infty}T(n_1,n_2)
   \Bigg|\mathcal{L}_{\frac{|n_1+n_2|}{2}}^{\frac{|n_1-n_2|}{2}}(F_3)\Bigg|^2\ln \left(NT(n_1,n_2)
   \Bigg|\mathcal{L}_{\frac{|n_1+n_2|}{2}}^{\frac{|n_1-n_2|}{2}}(F_3)\Bigg|^2\right)\\\nonumber
   \geq &-&\sum\limits_{n_1,n_2=0}^{\infty}T(n_1,n_2)
    \Bigg|\mathcal{L}_{\frac{|n_1+n_2|}{2}}^{\frac{|n_1-n_2|}{2}}(F_3)\Bigg|^2\ln \left(NT(n_1,n_2)
    \Bigg|\mathcal{L}_{\frac{|n_1+n_2|}{2}}^{\frac{|n_1-n_2|}{2}}(F_3)\Bigg|^2\right),
  \end{eqnarray}
  where we use the notation
   \begin{eqnarray*}T(n_1,n_2)&=&\exp\left(-\Bigg|\ln(\frac{n_1!}{n_2!})\Bigg|\right)F_1^{\frac{n_1-n_2}{2}}F_2^{\frac{n_1+n_2}{2}}.
  \end{eqnarray*}
\par  We deduce three new inequalities \eqref{2}, \eqref{28}, \eqref{11} for the classical special functions.
 Needless to say that the latter method gives us the opportunity to construct more different inequalitites for the classical polinomials. To this end, we can partition the matrix \eqref{13} into block matrices of any size, e.g. $3\times 3$. The two new matrices can be constructed by the following rule
 \begin{eqnarray*}
\widetilde{\rho}_{1}&=&\left(
              \begin{array}{cccc}
                P_0 + P_1+P_2& 0 & 0 & \cdots \\
                0& P_3 +P_4+P_5& 0& \cdots \\
                0 & 0 & P_6 +P_7+P_8& \cdots \\
                \vdots & \vdots& \vdots & \ddots \\
              \end{array}
            \right),
\end{eqnarray*}
 \begin{eqnarray*}
\widetilde{\rho}_{2}&=&
\left(
  \begin{array}{ccc}
    P_0 + P_3+P_6 +\ldots& 0&0 \\
    0 & P_1 + P_4+P_7 +\ldots&0 \\
      0 &0 &P_2 + P_5+P_8 +\ldots \\
  \end{array}
\right).
\end{eqnarray*}
The Shannon entropies of the subsystems with the latter matrices are
\begin{eqnarray}\label{14}
\widetilde{\mathcal{H}}(1)&=&-\sum\limits_{k=0}^{\infty}(P_{3k}+P_{3k+1}+P_{3k+2})\ln (P_{3k}+P_{3k+1}+P_{3k+2}),\\\nonumber
\widetilde{\mathcal{H}}(2)&=&-\left(\sum\limits_{k=0}^{\infty}P_{3k}\right)\ln \left(\sum\limits_{k=0}^{\infty}P_{3k}\right)
-\left(\sum\limits_{k=0}^{\infty}P_{3k+1}\right)\ln \left(\sum\limits_{k=0}^{\infty}P_{3k+1}\right)\\\nonumber
&-&\left(\sum\limits_{k=0}^{\infty}P_{3k+2}\right)\ln \left(\sum\limits_{k=0}^{\infty}P_{3k+2}\right)
\end{eqnarray}
and the Shannon information is the following
\begin{eqnarray}\label{19}\widetilde{I}&=&\widetilde{\mathcal{H}}(1)+\widetilde{\mathcal{H}}(2)-\mathcal{H}(12).
\end{eqnarray}
We can use it  to construct other inequalities for the classical special functions. The latter method provides an opportunity to obtain many other inequalities with the special structures and conditions for polynomials.

  \section{The Quadrature Uncertainty Relation and the Distribution Function}\label{sec:4}

  Let us select the dispersion matrix \eqref{34} as follows
 \begin{eqnarray}\label{31}\Sigma(x,y,t)&=&\left(
                             \begin{array}{cc}
                               x & t \\
                               t& y\\
                             \end{array}
                           \right).
  \end{eqnarray}
The invariant parameters of the latter matrix  are $Tr \Sigma=x+y$ and $\det{\Sigma}=xy-t^2$. Let $<q>=<p>=0$.
 The matrix $R$ has the following elements
   \begin{eqnarray*}&&R_{11}=R_{22}^{*}=-\frac{y-x+2ti}{x-2t^2+y+2xy+1/2},\quad
   R_{12}=\frac{2t^2-2xy+1/2}{x-2t^2+y+2xy+1/2}
  \end{eqnarray*}
    and the arguments of the Hermite  polynomial \eqref{1} are  $y_1=y_2^{*}=0$.
  The probability to have no photon is
  \begin{eqnarray*}P_0=2\left(2x-4t^2+2y+4xy+1\right)^{-\frac{1}{2}}.
  \end{eqnarray*}
Thus, the photon distribution function \eqref{1} can be rewritten as
    \begin{eqnarray}\label{29}\!\!P_n=2n!\sum\limits_{k=0}^{n}\!\!\frac{(-1)^k|H_{n-k}(0)|^2}{k!((n-k)!)^2}\frac{(-4(xy-t^2)+1)^k(-4(xy-t^2)+(x+y)^2)^{\frac{n-k}{2}}}{(4(xy-t^2)+2(x+y)+1)^{n+\frac{1}{2}}}.
  \end{eqnarray}
It is known, that the elements of the matrix $\Sigma(x,y,t)$ have to satisfy the inequality
 \begin{eqnarray}xy-t^2\geq1/4\label{30}
  \end{eqnarray}
  called the quadrature uncertainty relation \cite{Heisenberg,Robertson,schredinger:35,Dodonov2}.
  Let the latter inequality be violated, i.e. $xy-t^2=1/4-\tau$, $\tau\geq0$.
   Therefore, we rewrite the probability \eqref{29} with respect to the latter condition as
   \begin{eqnarray*}P_n(\tau,x,y)&=&2^n(-1)^nn!\sum\limits_{i=0}^{[n/2]}\frac{\tau^{n-2i}}{i!((n-2i)!)^2}\frac{((x+y)^2-1-4\tau)^{i}}{(x+y+1-4\tau)^{n+\frac{1}{2}}}.
  \end{eqnarray*}
  It is evident that $n=2l$, $l=0,1,\ldots$. Hence, we can write
     \begin{eqnarray*}&&P_{2l}(\tau, x,y,t)=(2l)!\sum\limits_{i=0}^{l}\frac{\tau^{2(l-i)}}{i!((2(l-i))!)^2}\frac{((x+y)^2-1-4\tau)^{i}}{2^{4i-2l-1/2}(x+y+1-4\tau)^{2l+\frac{1}{2}}}\\
     &=&(2l)!\sum\limits_{i=0}^{l}\frac{\tau^{2(l-i)}}{i!((2(l-i))!)^2}\frac{(\frac{1}{16}+t^4+\tau^2+\frac{t^2-\tau-y^2}{2}-2t^2\tau+y^4
     -6y^2\tau +2t^2y^2)^{i}}{2^{4i-2l-\frac{1}{2}}y^{2i-2l-\frac{1}{2}}(\frac{1}{4}-\tau+t^2+y^2+y-4y\tau)^{2l+\frac{1}{2}}}.
  \end{eqnarray*}
  Note that when the condition \eqref{30} is not satisfied, the latter function can become negative or complex and hence, it can not be the probability function.
  In the special case when $t=0$ we can rewrite the latter probability as
  \begin{eqnarray*}P_{2l}(\tau,y)&=&(2l)!\sum\limits_{i=0}^{l}\frac{\tau^{2(l-i)}}{i!((2(l-i))!)^2}\frac{(\frac{1}{16}+\tau^2-\frac{\tau+y^2}{2}+y^4
    -6y^2\tau)^{i}}{2^{4i-2l-\frac{1}{2}}y^{2i-2l-\frac{1}{2}}(\frac{1}{4}-\tau+y^2+y-4y\tau)^{2l+\frac{1}{2}}}
  \end{eqnarray*}
and the mean value is \begin{eqnarray*}\label{35}\langle n\rangle=-\det\left(\frac{R+\sigma_+}{R+\sigma_+}\right)=\frac{2(x-y)}{6x-2y+4xy+1}.\end{eqnarray*}
  As an example let us select the dispersion matrix \eqref{31} with the parameters $t=0$, $y=5$ and $\tau = 4$. The function $P_{2l}$ for such matrix is complex
    \begin{eqnarray}\label{33}P_{2l}&=&(2l)!\frac{2^{6l+\frac{1}{2}}5^{2l+\frac{1}{2}}}{\left(-\frac{215}{4}\right)^{2l+\frac{1}{2}}}\sum\limits_{k=0}^{l}\frac{\left(\frac{17}{4096}\right)^k}{k!((2(l-k))!)^2}
  \end{eqnarray}
  and the mean value is $\langle n\rangle=-23/57$.
  Using the definition of the complex logarithm function $\ln(z)=\ln(r)+i(\varphi+2\pi n)$, $z=re^{i(\varphi+2\pi n)}$, $n=0,\pm1,\pm2,\ldots$ and \eqref{9} we can define the entropy for such "probabilities" as
\begin{eqnarray}\label{32}\mathcal{H}_{-}(12)&=&-\sum\limits_{l=0}^{\infty}P_{2l}\left(\ln(|P_{2l}|)+i(\varphi+2\pi n)\right),\\\nonumber
\mathcal{H}_{-}(1)&=&-\sum\limits_{l=0}^{\infty}P_{2l}\left(\ln(|P_{2l}|)+i(\varphi+2\pi k)\right),\\\nonumber
\mathcal{H}_{-}(2)&=&-\left(\sum\limits_{l=0}^{\infty}P_{2l}\right)\left(\ln \left(\sum\limits_{l=0}^{\infty}|P_{2l}|\right)+i(\varphi+2\pi n)\right),
\end{eqnarray}
where $P_{2l}=ib\equiv|P_{2l}|e^{i(\varphi+2\pi n)}$, $\varphi=\pi/2$ if $b>0 $ and $\varphi=-\pi/2$ if $b>0 $. Hence, the information for such entropies can be defined as
 \begin{eqnarray*}I_{-}&=&\mathcal{H}_{-}(1)+\mathcal{H}_{-}(2)-\mathcal{H}_{-}(12)\\
 &=&-\left(\sum\limits_{l=0}^{\infty}|P_{2l}|e^{i(\varphi+2\pi n)}\right)\left(\ln \left(\sum\limits_{l=0}^{\infty}|P_{2l}|\right)+i(\varphi+2\pi n)\right).
  \end{eqnarray*}
  The latter information takes the complex values. For function \eqref{33} the information is
   \begin{eqnarray*}I_{-}&=&e^{i(\varphi+2\pi n)}\left(\sum\limits_{l=0}^{\infty}\frac{2^{8l+1}}{43^{2l+\frac{1}{2}}}\sum\limits_{k=0}^{l}\frac{(2l)!\left(\frac{17}{4096}\right)^k}{k!((2(l-k))!)^2}\right)\\
   &\cdot&\left(\ln \left(\sum\limits_{l=0}^{\infty}\frac{2^{8l+1}}{43^{2l+\frac{1}{2}}}\sum\limits_{k=0}^{l}\frac{(2l)!\left(\frac{17}{4096}\right)^k}{k!((2(l-k))!)^2}\right)+i(\varphi+2\pi n)\right)=0.
  \end{eqnarray*}
  However, we can also use \eqref{14} to define other entropies and the information for the latter "probabilities".

 \section{Examples}\label{sec:5}
\begin{figure}[htbp]
  \begin{center}
  \begin{minipage}[ht]{0.49\linewidth}
\centerline{\psfig{file=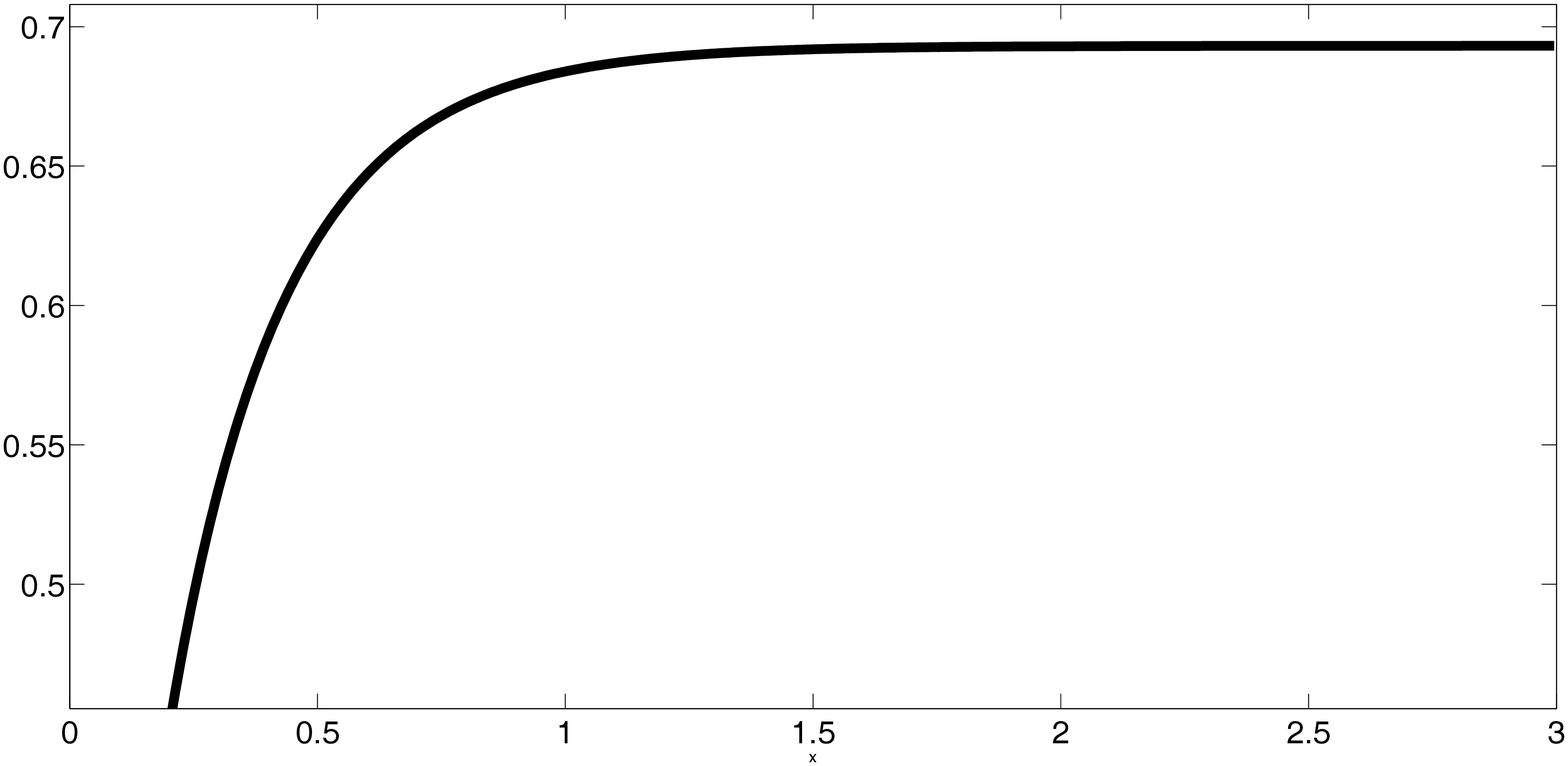, width=6.5cm}}
\vspace*{8pt}
\caption{The information \eqref{7} for various values of $\overline{x}$.\label{fig:1}}
\end{minipage}
\begin{minipage}[ht]{0.49\linewidth}
\centerline{\psfig{file=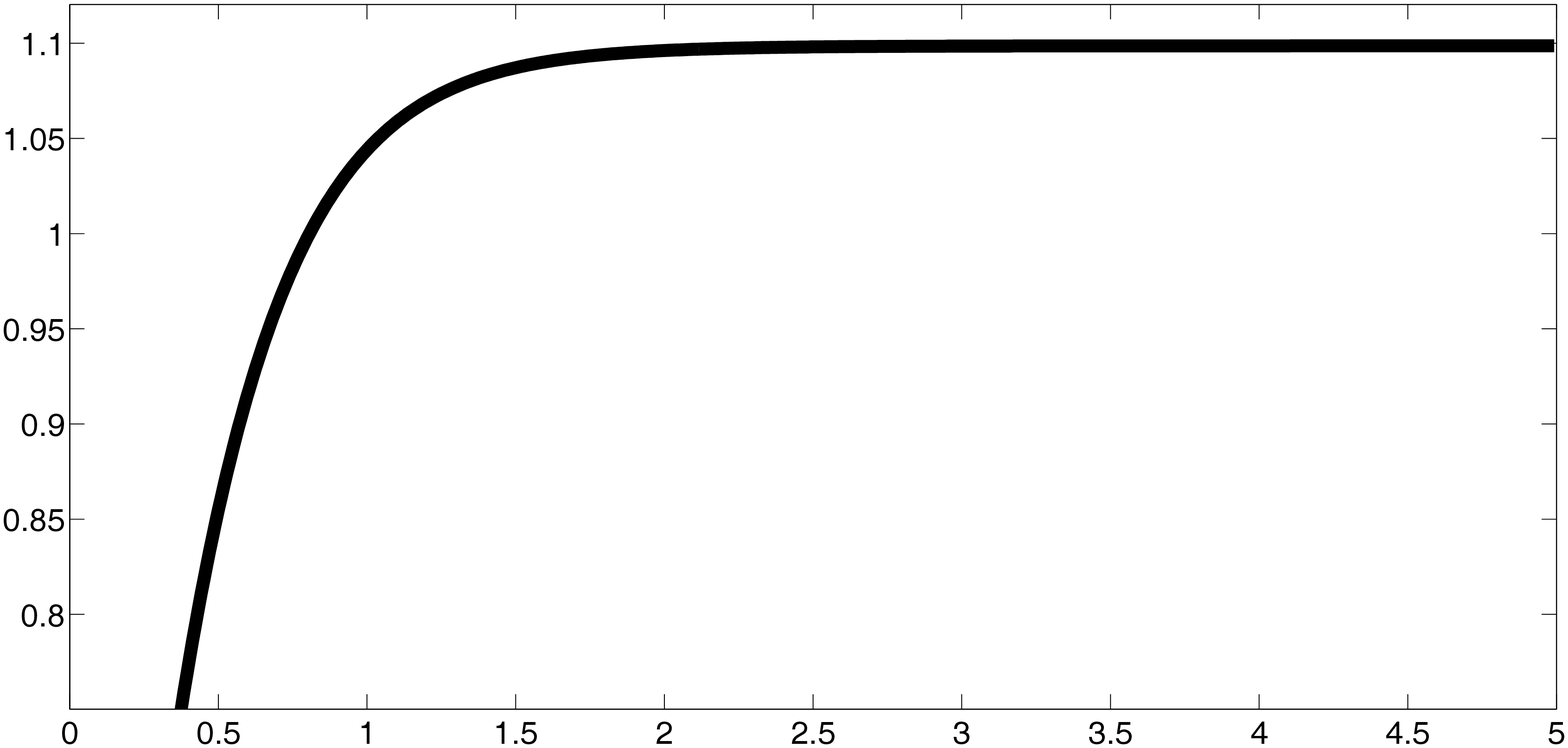, width=6.5cm}}
\vspace*{8pt}
\caption{The information \eqref{19} for various values of $\overline{x}$.\label{fig:1_1}}
\end{minipage}
\end{center}
\end{figure}

     As a first example we select the coherent state $|\alpha\rangle$, which is the eigenstate of the annihilation operator $\widehat{a}$ associated to the eigenvalue $\alpha$,
     i.e. $\widehat{a}|\alpha\rangle=\alpha|\alpha\rangle$. It is known, that $\alpha\in\mathbb{C}$ can be represented as $\alpha = |\alpha|e^{i\theta}$,
where $|\alpha|$ and  $\theta$ are real numbers called the amplitude and the phase of the state, respectively. The coherent state can be dicomposed in the basis of Fock states
        \begin{eqnarray*}|\alpha\rangle =e^{-{|\alpha|^2\over2}}\sum_{n=0}^{\infty}{\alpha^n\over\sqrt{n!}}|n\rangle =e^{-{|\alpha|^2\over2}}e^{\alpha\hat a^\dagger}|0\rangle,  \end{eqnarray*}
where  $|n\rangle$ are  the photon numbers and the eigenvectors of the Hamiltonian
       \begin{eqnarray*}\mathcal{H} = \hat a^\dagger \hat a + 1/2.\end{eqnarray*}
     One can see, that  the corresponding Poissonian distribution
        \begin{eqnarray*}P(n)&=&e^{-\langle n\rangle}\frac{\langle n\rangle^n}{n!},
  \end{eqnarray*}
is the probability of detecting $n$ photons with the mean photon number $<n>=|\alpha|^2$ and the dispersion $(\triangle n)^2=|\alpha|^2$.
    Due to this, we study the Poissonian  distribution function
   \begin{eqnarray*}P_n(\overline{x})&=&e^{-\overline{x}}\frac{\overline{x}^n}{n!}.
  \end{eqnarray*}
Using the latter distribution and
  since $\lim\limits_{n\rightarrow\infty}{e^{-\overline{x}}\frac{\overline{x}^n}{n!}}=0$, $\lim\limits_{x\rightarrow 0}{x\ln{x}}=0$ the Shannon information \eqref{10} for the Poissonian  distribution function is
     \begin{eqnarray}\label{7}I=&-&\left(\sum\limits_{k=0}^{\infty}e^{-\overline{x}}\frac{\overline{x}^{2k}}{(2k)!}\right)\ln \left(\sum\limits_{k=0}^{\infty}e^{-\overline{x}}\frac{\overline{x}^{2k}}{(2k)!}\right)\\\nonumber
&-&\left(\sum\limits_{k=0}^{\infty}e^{-\overline{x}}\frac{\overline{x}^{2k+1}}{(2k+1)!}\right)\ln \left(\sum\limits_{k=0}^{\infty}e^{-\overline{x}}\frac{\overline{x}^{2k+1}}{(2k+1)!}\right)\\\nonumber
&=&-e^{-\overline{x}}\left(\sinh{\overline{x}}\ln\left(e^{-\overline{x}}\sinh{\overline{x}}\right)+\cosh{\overline{x}}\ln\left(e^{-\overline{x}}\cosh{\overline{x}}\right)\right).
  \end{eqnarray}
  The information \eqref{7} is shown in Figure \ref{fig:1} for various values of $\overline{x}$.
However,  information \eqref{19} for the Poissonian  distribution function is different and has the following form
   \begin{eqnarray}\label{20}\widetilde{I}&=&-\frac{1}{3}e^{-\frac{3x}{2}}\left(e^{\frac{3x}{2}}-2\sin\left(\frac{\pi-3\sqrt{3}x}{6}\right)\right)\ln\left(\frac{1}{3} - \frac{2}{3}e^{-\frac{3x}{2}}\sin\left(\frac{\pi-3\sqrt{3}x}{6}\right)\right)\nonumber\\
   &-&\frac{1}{3}\left(2e^{-\frac{3x}{2}}\cos\left(\frac{\sqrt{3}x}{2}\right)+1\right)\ln\left(\frac{1}{3}\left(2e^{-\frac{3x}{2}}\cos\left(\frac{\sqrt{3}x}{2}\right)+1\right)\right)\\\nonumber
   &-&\frac{1}{3}e^{-\frac{3x}{2}}\left(e^{\frac{3x}{2}}-2\sin\left(\frac{\pi-3\sqrt{3}x}{6}\right)\right)\ln\left(\frac{1}{3} - \frac{2}{3}e^{-\frac{3x}{2}}\sin\left(\frac{\pi+3\sqrt{3}x}{6}\right)\right).
  \end{eqnarray}
  Information \eqref{20} is shown in  Figure \ref{fig:1_1} for various values of $\overline{x}$. We can see, that using different mappings we can get complitely different informations.
 \par
\begin{figure}[htbp]
\centerline{\psfig{file=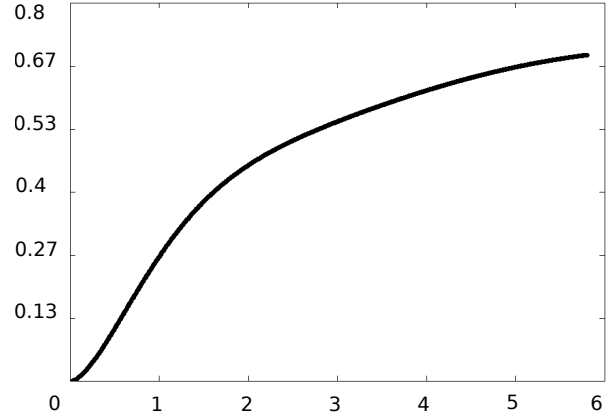, width=8cm}}
\vspace*{8pt}
\caption{The information \eqref{37} for $\lambda=2$  and  various values of $\alpha$.\label{fig:3}}
\end{figure}

 In the second example we consider nonlinear dynamical systems related to deformations of
linear classical and quantum systems like the nonlinear coherent state. To produce from the linear system the nonlinear one the  parameters of the linear system  are replaced with constants of the motion of the nonlinear system \cite{Aric,Daskaloyannis}. For the  $q$-oscillators \cite{Biedenharn,Macfarlane} the frequency of vibrations was replaced by the
constant of the motion that depend on the amplitude of the
vibrations. Thus, the $q$-oscillator is a specific nonlinear physical system. Some experimental results for the $q$-oscillator can be found in \cite{Angelis,Hilborn,Tino}.
 \\We consider the eigenfunctions of $A$, $|\alpha,f\rangle$ in the Hilbert space,
  i.e. $A|\alpha,f\rangle=\alpha|\alpha,f\rangle$, $\alpha\in\mathbb{C}$. The latter state can be dicomposed in the basis of the Fock space as
        \begin{eqnarray*}|\alpha,f\rangle =\sum_{n=0}^{\infty}{c_0\alpha^n\over\sqrt{n!}(f(n)!)^2}|n\rangle,\end{eqnarray*}
  where $f(n)!=f(0)f(1)\cdots f(n)$ and
  \begin{eqnarray*}c_0=\left(\sum_{n=0}^{\infty}{|\alpha|^{2n}\over\sqrt{n!}(f(n)!)^2}\right)^{-1/2}.\end{eqnarray*}
  The corresponding photon distribution in the $f$-coherent state is
        \begin{eqnarray*}P_{n,f(n)}(\alpha)&=&\left(\sum\limits_{j=0}^{\infty}{|\alpha|^{2j}\over\sqrt{j!}(f(j)!)^2}\right)^{-1}{|\alpha|^{2n}\over\sqrt{n!}(f(n)!)^2}=C_0{|\alpha|^{2n}\over\sqrt{n!}(f(n)!)^2}.
  \end{eqnarray*}
  Substituting $P_{n,f(n)}(\alpha)$ in information inequality \eqref{27} we can write
  \begin{eqnarray}\label{36}&-&\left(\sum\limits_{k=0}^{\infty}{|\alpha|^{4k}\over\sqrt{2k!}(f(2k)!)^2}\right)\ln \left(\sum\limits_{k=0}^{\infty}{C_0|\alpha|^{4k}\over\sqrt{2k!}(f(2k)!)^2}\right)\\\nonumber
&-&\left(\sum\limits_{k=0}^{\infty} {|\alpha|^{2(2k+1)}\over\sqrt{(2k+1)!}(f(2k+1)!)^2}\right)\ln \left(\sum\limits_{k=0}^{\infty}  {C_0|\alpha|^{2(2k+1)}\over\sqrt{(2k+1)!}(f(2k+1)!)^2}\right)\\\nonumber
&-&\sum\limits_{k=0}^{\infty}({|\alpha|^{4k}\over\sqrt{2k!}(f(2k)!)^2}+{|\alpha|^{2(2k+1)}\over\sqrt{(2k+1)!}(f(2k+1)!)^2})\\\nonumber
&\cdot&\ln ({C_0|\alpha|^{4k}\over\sqrt{2k!}(f(2k)!)^2}+  {C_0|\alpha|^{2(2k+1)}\over\sqrt{(2k+1)!}(f(2k+1)!)^2})\\\nonumber
&\geq& -\sum\limits_{k=0}^{\infty}{|\alpha|^{2k}\over\sqrt{k!}(f(k)!)^2}\ln   {C_0|\alpha|^{2k}\over\sqrt{k!}(f(k)!)^2}.
\end{eqnarray}
Let us take the photon distribution in the $q$-coherent state
        \begin{eqnarray*}P_{\lambda,\alpha}(n)&=&\left(\sum\limits_{j=0}^{\infty}{|\alpha|^{2j}\over\left(\frac{\sinh\lambda j}{\sinh\lambda }\right)!}\right)^{-1}{{|\alpha|^{2n}\over\left(\frac{\sinh\lambda n}{\sinh\lambda }\right)!}}=c_0{{|\alpha|^{2n}\over\left(\frac{\sinh\lambda n}{\sinh\lambda }\right)!}}.
  \end{eqnarray*}
Since $\lim\limits_{n\rightarrow\infty}{{|\alpha|^{2j}\over\left(\frac{\sinh\lambda n}{\sinh\lambda }\right)!}}=0$ if $n\gg1/\lambda$, $\lim\limits_{x\rightarrow 0}{x\ln{x}}=0$  inequality \eqref{36} can be rewritten as
  \begin{eqnarray}\label{37}I&=&-\left(\sum\limits_{k=0}^{\infty}c_0{{|\alpha|^{4k}\over\left(\frac{\sinh\lambda 2k}{\sinh\lambda }\right)!}}\right)\ln \left(\sum\limits_{k=0}^{\infty}c_0{{|\alpha|^{4k}\over\left(\frac{\sinh\lambda 2k}{\sinh\lambda }\right)!}}\right)\\\nonumber
&-&\left(\sum\limits_{k=0}^{\infty} c_0{{|\alpha|^{2(2k+1)}\over\left(\frac{\sinh\lambda (2k+1)}{\sinh\lambda }\right)!}}\right)\ln \left(\sum\limits_{k=0}^{\infty}  c_0{{|\alpha|^{2(2k+1)}\over\left(\frac{\sinh\lambda (2k+1)}{\sinh\lambda }\right)!}}\right)\geq 0.
\end{eqnarray}
For example, for $\lambda=2$ information \eqref{37} is shown in Figure \ref{fig:3} for  various values of $\alpha$.
  \par
\begin{figure}[htbp]
\centerline{\psfig{file=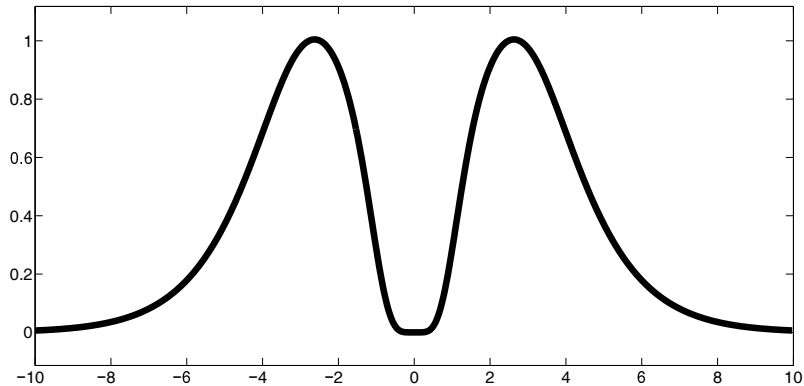, width=13cm}}
\vspace*{8pt}
\caption{The information \eqref{19} for the special case of the squeezed vacuum state \eqref{12} for the various values of $r$.\label{fig:2}}
\end{figure}

 The third example concerns the squeezed and correlated state. In this state the dispersion matrix is
   \begin{eqnarray*}\Sigma&=&\frac{1}{2}\left(
                             \begin{array}{cc}
                              \cosh{2r}+\cos\theta\sinh{2r}& \sin\theta\sinh{2r} \\
                               \sin\theta\sinh{2r} & \cosh{2r}-\cos\theta\sinh{2r} \\
                             \end{array}
                           \right),
  \end{eqnarray*}
  where $r$ and $\theta$ determine the dispersions of the quadrature components.
  The  photon distribution for the squeezed light is
 \begin{eqnarray*}P_n &=&P_0\frac{(\tanh{r})^n}{n!2^n}\Bigg|H_n(g(p,q))\Bigg|^2\equiv\mathrm{ H}_{n},\\
 g(p,q)&\equiv&e^{-\frac{i\theta}{2}}\sqrt{\tanh{r}}\left(\frac{<q>-i<p>}{2}+e^{i\theta}\coth r\frac{<q>+i<p>}{2}\right),
  \end{eqnarray*}
  where the probability to have no photon is
     \begin{eqnarray*}P_0&=&\frac{1}{\cosh{r}}\exp\left(-\frac{<p>^2+<q>^2}{2}\right)\\
     &+&\frac{\tanh{r}}{2}\left((<p>^2-<q>^2)\cos{\theta}+2<p><q>\sin{\theta}\right).
  \end{eqnarray*}
    Substituting the latter distribution in \eqref{27} we can write the following inequalitity
    \begin{eqnarray*}&-&\left(\sum\limits_{k=0}^{\infty}\mathrm{H}_{2k}\right)\ln \left(\sum\limits_{k=0}^{\infty}\mathrm{H}_{2k}\right)-\left(\sum\limits_{k=0}^{\infty}\mathrm{H}_{2k+1}\right)\ln \left(\sum\limits_{k=0}^{\infty}\mathrm{H}_{2k+1}\right)\\\nonumber
&-&\sum\limits_{k=0}^{\infty}\left(\mathrm{H}_{2k}+\mathrm{H}_{2k+1}\right)\ln \left(\mathrm{H}_{2k}+\mathrm{H}_{2k+1}\right)\geq -\sum\limits_{k=0}^{\infty}\mathrm{H}_{k}\ln \mathrm{H}_{k}.
\end{eqnarray*}
For the special case of the squeezed vacuum state $<p>=<q>=0$ and for $\theta=0$ the photon distribution is reducted to
\begin{eqnarray}\label{12}P_{2n}&=&\frac{1}{\cosh{r}}\left(\frac{\tanh{r}}{2}\right)^{2n}\frac{2n!}{(n!)^2},\quad P_{2n+1}=0
\end{eqnarray}
and the Shannon information is reduced to
\begin{eqnarray*}\label{8}I=&-&\left(\sum\limits_{k=0}^{\infty}P_{2k}\right)\ln \left(\sum\limits_{k=0}^{\infty}P_{2k}\right)= -\cosh{r}\cdot sech{r}\ln(\cosh{r}\cdot sech{r})=0.
\end{eqnarray*}
On the other hand,  Shannon entropies \eqref{14} for the latter probabilities are
\begin{eqnarray*}
\widetilde{\mathcal{H}}(1)&=&-\sum\limits_{k=0}^{\infty}\frac{1}{\cosh{r}}\left(\frac{\tanh{r}}{2}\right)^{6k}
\Bigg(\frac{6k!}{(3k!)^2}+\left(\frac{\tanh{r}}{2}\right)^{2}\frac{(6k+2)!}{((3k+1)!)^2}\\\nonumber
&+&\left(\frac{\tanh{r}}{2}\right)^{4}\frac{(6k+4)!}{((3k+2)!)^2}\Bigg)\ln \Bigg(\frac{1}{\cosh{r}}\left(\frac{\tanh{r}}{2}\right)^{6k}
\Bigg(\frac{6k!}{(3k!)^2}\\\nonumber
&+&\left(\frac{\tanh{r}}{2}\right)^{2}\frac{(6k+2)!}{((3k+1)!)^2}+\left(\frac{\tanh{r}}{2}\right)^{4}\frac{(6k+4)!}{((3k+2)!)^2}\Bigg)\Bigg),
\end{eqnarray*}\begin{eqnarray*}
\widetilde{\mathcal{H}}(2)&=&-\left(\sum\limits_{k=0}^{\infty}\frac{1}{\cosh{r}}\left(\frac{\tanh{r}}{2}\right)^{6k}\frac{6k!}{(3k!)^2}\right)\ln \left(\sum\limits_{k=0}^{\infty}\frac{1}{\cosh{r}}\left(\frac{\tanh{r}}{2}\right)^{6k}\frac{6k!}{(3k!)^2}\right)\\\nonumber
&-&\left(\sum\limits_{k=0}^{\infty}\frac{1}{\cosh{r}}\left(\frac{\tanh{r}}{2}\right)^{6k+4}\frac{(6k+4)!}{((3k+2)!)^2}\right)\\\nonumber
&\cdot&\ln \left(\sum\limits_{k=0}^{\infty}\frac{1}{\cosh{r}}\left(\frac{\tanh{r}}{2}\right)^{6k+4}\frac{(6k+4)!}{((3k+2)!)^2}\right)\\\nonumber
&-&\left(\sum\limits_{k=0}^{\infty}\frac{1}{\cosh{r}}\left(\frac{\tanh{r}}{2}\right)^{6k+2}\frac{(6k+2)!}{((3k+1)!)^2}\right)\\\nonumber
&\cdot&\ln \left(\sum\limits_{k=0}^{\infty}\frac{1}{\cosh{r}}\left(\frac{\tanh{r}}{2}\right)^{6k+2}\frac{(6k+2)!}{((3k+1)!)^2}\right).
\end{eqnarray*}
Using the latter entropies and since \begin{eqnarray*}&&\lim\limits_{k\rightarrow\infty}\{\left(\frac{\tanh{r}}{2}\right)^{6k}
\Bigg(\frac{6k!}{(3k!)^2}+\left(\frac{\tanh{r}}{2}\right)^{2}\frac{(6k+2)!}{((3k+1)!)^2}\\
&+&\left(\frac{\tanh{r}}{2}\right)^{4}\frac{(6k+4)!}{((3k+2)!)^2}\Bigg)\}=0,\quad \lim\limits_{x\rightarrow 0}{x\ln{x}}=0\end{eqnarray*}
hold, we can obtain the information \eqref{19}. The result is shown in Figure \ref{fig:2} for various values of $\overline{r}$.
Hence, we illustrate that different mappings may provide different kinds of inequalities for the special functions.

\section{Conclusions}
  To conclude we point out the main results of our work. Considering the probability function polynomial representation and applying the known subadditivity condition for joint probability distributions the new inequalities for the Hermite, the Legendre and the Laguerre polynomials
  are obtained. The inequalities correspond to the entropic inequalities for the Shannon entropies of the bipartite systems. The results are shown in detail on the example of the Poissonian  distribution function and for the special case of the squeezed vacuum state and the q-cocherent state, where the Shannon  information of the bipartite system is expressed in terms of the polynomials. The dependence between the  quadrature uncertainty relation and the existence of the photon probability distribution is shown. Our inferences are that the violation of the quadrature uncertainty relation leads to the fact that the distribution function can take negative or even complex values. In other words, the function is no longer the probability.
\section*{Acknowledgements}
Markovich L.A. was partly supported by the Russian Foundation for Basic Research, grant 16-08-01285 A.


\begin{thebibliography}{0}

\bibitem{Dodonov}V.V. Dodonov, O.V. Man'ko, and V.I. Man'ko, {\it Photon distribution for one mode mixed light with generic gaussian Wigner function},  {\sl Phys. Rev. A}, \textbf{ 49}, 2993, 1994.
\bibitem{Schleich}G. Schrade, V.M. Akulin, V.I. Man'ko, and W. P. Schleich, {\it Photon statistics of a two-mode squeezed vacuum},  {\sl Phys. Rev. A}, \textbf{48}, 2398, 1993.
\bibitem{Lieb}E.H. Lieb, M.B. Ruskai, {\it Proof of the Strong Subadditivity of Quantum Mechanical Entropy}, {\sl J. Math. Phys.}, \textbf{14}, 1938--1941, 1973.
\bibitem{Chernega}V.N. Chernega and V.I. Man'ko, {\sl J. Russ. Laser Res.}, \textbf{29}, 505, 2008.
\bibitem{Mendes}M.A. Man'ko, V.I. Man'ko, and R. Vilela Mendes, {\sl J. Russ. Laser Res.}, \textbf{27}, 507, 2006.
\bibitem{Mar2}V.I. Man'ko and L.A. Markovich, {\it  New inequalities for quantum  von Neumann and tomographic mutual information}, {\sl  J. Russ. Laser Res.}, \textbf{35(4)}, 355--361, (2014).
\bibitem{Mar0}V.I. Man'ko, L.A. Markovich, {\it  Entropic inequalities and properties of some special functions},  {\sl J. Russ. Laser Res.}, \textbf{35(2)}, 200--210, 2014.
\bibitem{Mar10}V.I. Man'ko, L.A. Markovich, {\it  Entropic inequalities for matrix elements of rotation group irreducible
representations},  {\sl  arXiv:1511.07341}, 2015.
 \bibitem{Ibort} A. Ibort,  V.I. Man'ko, G. Marmo, A. Simoni, F. Ventriglia, {\it An introduction to the tomographic picture of quantum mechanics}, Physica Scripta \textbf{79(6)}, 2009.
     \bibitem{Marian} P. Marian, T.A. Marian, {\it Squeezed states with thermal noise. I. Photon-number statistics},  {\sl Phys. Rev. A},  \textbf{47}, 4474, 1993.
\bibitem{Marian2} P. Marian, T.A. Marian, {\it    Squeezed states with thermal noise. II. Damping and photon counting},  {\sl Phys. Rev. A},  \textbf{47}, 4487, 1993.
\bibitem{Olivares} S. Olivares, {\it Quantum optics in the phase space. A tutorial on Gaussian states},  {\sl The European Physical Journal Special Topics}, \textbf{ 203(1)}, 3--24, 2012.
    \bibitem{Heisenberg} W. Heisenberg, {\it \"{U}ber den anschaulichen Inhalt der quantentheoretischen Kinematik und Mechanik},  {\sl Zeitschrift f\"{u}r Physik},  \textbf{43(3)}, 172--198, 1927.
\bibitem{Robertson} H.P. Robertson, {\it The Uncertainty Principle},  {\sl Phys. Rev.},  \textbf{34}, 163, 1929.
\bibitem{schredinger:35}E. Schr\"{o}dinger, {\it Zum Heisenbergschen Unscharfeprinzip
Sitzungsberichte der Preusischen Akademie der Wissenschaften},  {\sl Physikalisch-mathematische Klasse}, 296--303, 1930.
\bibitem{Dodonov2}   V. Dodonov,  V.I. Man'ko, {\it Generalization of uncertainty relation in quantum mechanics},  {\sl
Proc. P. N. Lebedev Physical Institute (Trudy FIAN)},  \textbf{183} Markov M.A., ed (New York: NOVA), 3--101, 1989.
\bibitem{Schrade} O.V. Manko, G. Schrade, {\it Photon statistics of $2$-mode squeezed-light with Gaussian-Wigner function},  {\sl Physica scripta. T}, \textbf{58(3)}, 228--234, 1998.
\bibitem{Shannon}C.E. Shannon,  {\it A Mathematical Theory of Communication}, {\sl Bell System Technical Journal}, \textbf{27}, 379, 1948.
\bibitem{Lopez}R. López-Peña, V.I. Man'ko, G. Marmo, E.C.G. Sudarshan, F. Zaccaria,  {\it Photon distribution in nonlinear coherent states},  {\sl J. Russ. Laser Res.}, \textbf{21(4)}, 305--316, 2000.
 \bibitem{Aric} M. Aric, D.D. Coon, Y. Lam, {\it   Introduction of a finite set of quon operators in the context of the dual resonnace model},  {\sl J. Math. Phys.},  \textbf{16}, 1776, 1975.
\bibitem{Daskaloyannis} C. Daskaloyannis, {\it Generalized Deformed Oscillator and Nonlinear Algebras},  {\sl J. Phys. A: Math. Gen.},  \textbf{
24(15)}, L789-L794, 1991.
\bibitem{Biedenharn} L.C. Biedenharn, {\it  The  quantum group $SU(2)_q$ and a  $q$-analogue of the boson operators},  {\sl J. Phys. A: Math. Gen.},  \textbf{22}, L873, 1989.
\bibitem{Macfarlane}A.J. Macfarlane,  {\it On $q$-analogues of the quantum harmonic oscillator and the quantum group $SU(2)_q$},  {\sl J. Phys. A: Math. Gen.}, \textbf{22} 4581, 1989.
\bibitem{Angelis}   M. de Angelis, G.Gagliardi, L. Gianfrani, and G. M. Tino,  {\it Test of the symmetrization
postulate for spin-$0$ particles},  {\sl Phys. Rev. Lett.},  \textbf{ 76}, 2840, 1996.
\bibitem{Hilborn} R.C. Hilborn, C.L. Yuca,  {\it Spectroscopic test of the symmetrization postulate for spin-$0$
nuclei},  {\sl Phys. Rev. Lett.},  \textbf{ 76}, 2844, 1996.
\bibitem{Tino} V.I. Man'ko, G.M. Tino, {\it Experimental limit of the blue-shift of the frquency of light
implied by q-nonlinearity},  {\sl Phys. Lett. A},  \textbf{202}, 24, 1995.

\end{thebibliography}
\end{document}